\documentstyle[12pt]{article}
\begin{document}
\begin{titlepage}
\renewcommand{\thefootnote}{\fnsymbol{footnote}}
\begin{center}
{\Large{\bf A Toy Model Approach to the Canonical Non-perturbative
Quantization of the Spatially Flat Robertson-Walker Spacetimes with
Cosmological Constant}
\vspace{20mm}\\}
{\large C. DARIESCU \footnote{MONBUSHO fellow, on leave of absence
from The Dept. of Theoretical Physics, ^^ ^^ Al.I. Cuza'' University,
6600 Ia\c{s}i, Romania}, S. HAMAMOTO \footnote{E-mail address:
mah00024@niftyserve.or.jp}} \\ and \\ {\large Marina-Aura DARIESCU
\footnote{On leave of absence from The Dept. of Theoretical Physics,
^^ ^^ Al.I. Cuza'' University, 6600 Ia\c{s}i, Romania}}
\vspace{10mm}\\
{\it Department of Physics \\ Toyama University \\ Gofuku, Toyama
930, Japan}
\end{center}
\vspace{20mm}
Short title: {\large{\bf{\it Quantization of the Robertson-Walker
Spacetime}}}
\vspace{10mm}\\
PACS numbers: 98.80.Hw, 04.60.Ds, 98.80.Cq
\end{titlepage}
\newpage
\thispagestyle{empty}
\begin{abstract}
We present a toy model approach to the canonical non-perturbative
quantization of the spatially-flat Robertson-Walker Universes with
cosmological constant, based on the fact that such models are exactly
solvable within the framework of a simple Lagrangian formulation. The
essential quantum dynamical metric-field and the corresponding
Hamiltonian, explicitly derived in terms of annihilation and creation
operators, point out that the Wheeler - DeWitt equation is a natural
(quantum) generalization of the $G_{44}$ - Einstein equation for the
classical De Sitter spacetime and selects the physical states of the
quantum De Sitter Universe. As a result of the exponential universal
expansion, the usual Fock states (defined as the eigenstates of the
number-operator) are no longer invariant under the derived
Hamiltonian. They exhibit quantum fluctuation of the energy and of
the metric field which lead to a (geometrical) volume quantization.
\end{abstract}

\newpage
\setcounter{page}{1}
\section{General}

The basic idea of this paper is to look for one of the most simple,
but physically interesting, spacetime structures with a minimum
number of essential dynamical variables, to write down the action
functional for the gravitational field in terms of these variables
and choosing the Lagrangian whose Euler-Lagrange equations have to be
compatible with the Einstein's equations for that spacetime, to
perform a canonical non-perturbative quantization. Is almost obvious
that such a spacetime might be the well known
Friedmann-Robertson-Walker Universe with cosmological constant and
flat $t$ - leveled hypersurfaces, described by the metric
\begin{equation}
ds^{2} = \eta_{ab} \omega^{a} \omega^{b} = e^{2f(t)}
\delta_{\alpha \beta} dx^{\alpha} dx^{\beta} - (dt)^{2}
\end{equation}
where
\begin{eqnarray}
a)  &  ( \eta_{ab} ) & =  {\rm diag}(1, \, 1, \, 1, \, -1)  \nonumber
\\
b) &  \omega^{\alpha} & =  e^{f(t)} dx^{\alpha}  \\
c) &  \omega^{4} & =   dt  \nonumber
\end{eqnarray}
and $f:R \rightarrow R$ is the primitive of the Hubble's function
$h = f,_{4}$ , where $( \cdot ),_{4} = \frac{\partial \cdot}
{\partial t}$ , since  the dual tetrads $\omega^{\alpha}$ are
entirely defined by $f(t)$ and the connection 1-forms (without
torsion)
\begin{equation}
\Gamma_{ab} \wedge \omega^{b} = - \eta_{ab} d \omega^{b}
\end{equation}
which, as can be easily checked, concretely read
\begin{eqnarray}
a) \; & \Gamma_{\alpha 4}  =  f,_{4} \omega^{\alpha} & \;   \nonumber
\\
&  & ,     \,  \alpha , \, \beta = \, \overline{1, \, 3} \\
b)  \; & \Gamma_{\alpha \beta}  =  0 & \;  \ \nonumber
\end{eqnarray}
are completely defined by the Hubble's function $f,_{4}$ . Due to the
general simplicity of the analysed system there is no need to engage
in full the
^^ ^^ 3 + 1 decomposition'' - formalism of quantization
\cite{Arn:Phy,Arn:Ann,Ast:Lec} because the metric (1), possesing the
scalar curvature
\begin{equation}
R = 6 \left[ f,_{44} + 2 (f,_{4})^2 \right] ,
\end{equation}
already leads to the following easily handling expression
\begin{equation}
S[f] = \frac{3}{\kappa_{0}} \int e^{3f} \left[f,_{44} + 2
(f,_{4} )^2 - \frac{\Lambda}{3} \right] \, d^{3} x \, dt \; \; , \;
\kappa_{0} = 8 \pi G
\end{equation}
for the action functional of the gravitational field in the spatially
flat FRW Universe with the cosmological constant $\Lambda$. Before
proceeding to quantization it is worthy to have a good image of the
classical dynamics of the system as it is described by Einstein's
equations. Sure, at this level (classically) nothing is new but when
the quantization is engaged it will turn out that the
$G_{44}$ - Einstein equation not only heavily lies on the Hamiltonian
associated to the Lagrangian derived from (6) but also directly
expresses the well known Wheeler-DeWitt equation \cite{DeW:Phy}
\begin{equation}
\hat H  \mid \psi \rangle = 0
\end{equation}
obeyed by the physical states $\mid \psi \rangle $ .

\section{Geometry and Exact Solutions}

The algebraically essential components of the Einstein tensor
$G_{ab}$ with respect to the dual pseudo-orthonormal basis (2) can be
worked out as usual, starting with the 2-nd Cartan structure equation
\begin{equation}
\Omega_{ab} = d \Gamma_{ab} + \Gamma_{ac} \wedge
\Gamma^{c}_{\cdot \, b}
\end{equation}
which leads to the curvature 2-forms
\begin{eqnarray}
& a) &  \Omega_{\alpha \beta}  =  (f,_{4})^{2} \omega^{\alpha} \wedge
\omega^{\beta} \nonumber \\
& &   \\
& b) &  \Omega_{\alpha 4}  =   - \left[ f,_{44} + (f,_{4})^{2}
\right] \omega^{\alpha} \wedge \omega^{4} \nonumber
\end{eqnarray}
from which the curvature components read
\begin{eqnarray}
& a) & R_{\alpha \beta \alpha \beta}  =  (f,_{4})^{2} \nonumber \\
& & \\ & b) & R_{\alpha 4 \alpha 4}  =  - \left[ f,_{44} +
(f,_{4})^{2} \right] \nonumber
\end{eqnarray}
(without summation after the repeated indices) and so, the Ricci
tensor components
\begin{equation}
R_{ab} = \eta^{cd} R_{cadb}
\end{equation}
get the expressions
\begin{eqnarray}
& a) & R_{\alpha \beta}  = \left[ f,_{44} + 3 (f,_{4})^{2} \right]
\delta_{\alpha \beta} \nonumber \\ & & \\ & b) & R_{44}  = - 3
\left[ f,_{44} + (f,_{4})^{2} \right] \nonumber
\end{eqnarray}
and the scalar curvature
\begin{equation}
R = \eta^{ab} R_{ab}
\end{equation}
is given by (5) and the only non-vanishing components of Einstein
tensor concretely read
\begin{eqnarray}
& a) & G_{\alpha \beta}  = - \left[ 2 f,_{44} + 3 (f,_{4})^{2}
\right] \delta_{\alpha \beta} \nonumber \\ & & \\
& b) & G_{44}  =  3  (f,_{4})^{2} \nonumber
\end{eqnarray}
Thus, the Einstein's equations with cosmological constant
\begin{equation}
G_{ab} + \eta_{ab} \Lambda = 0
\end{equation}
(in the empty spacetime case) reduce to the following overdetermined
system of two ordinary differential equations for the metric function
$f$:
\begin{eqnarray}
a) & 2 f,_{44} + 3 (f,_{4})^{2} & = \Lambda \nonumber \\ & &  \\
b) & 3 (f,_{4})^{2} & = \Lambda \nonumber
\end{eqnarray}
which are known respectively as the
^^ ^^ cosmological pressure'' - equation and the
^^ ^^ energy'' - equation. As $\Lambda$ is constant, is clear from
the ^^ ^^ energy'' - equation that
\begin{equation}
f,_{44} = 0
\end{equation}
and this assures the compatibility of (16), the ^^ ^^ cosmological
pressure'' - equation being actually the same with (16.b). Therefore,
classically, the dynamics of the spatially flat FRW Universe with
cosmological constant and no (ordinary) matter-content is entirely
encoded in the ^^ ^^ energy'' - equation (16.b) which posses the
general solutions
\begin{eqnarray}
& a) & f_{+} (t)  = f_{+}^{0} + \sqrt{\frac{\Lambda}{3}} \, t
\nonumber \\ & & \\ & b) &
f_{-} (t)  = f_{-}^{0} - \sqrt{\frac{\Lambda}{3}} \, t \nonumber
\end{eqnarray}
with $t \in ( - \infty , \, \infty)$ and $f_{+}^{0} , \, f_{-}^{0}
\in R$ being the scale factors. Setting $f_{\pm}^{0} = 0$ i.e. at
$t=0$ the analyzed spacetime is Minkowskian, the metric (1) becomes
\begin{equation}
ds^{2} = e^{\pm 2 \sqrt{\frac{\Lambda}{3}} t} \,
\delta_{\alpha \beta} dx^{\alpha} dx^{\beta} - \, (dt)^{2}
\end{equation}
and these two ^^ ^^ disjoint'' cosmological vacuum-type exact
solutions of Einstein's field equations are (very well) known as
describing the so called ^^ ^^ red-shifted'' and ^^ ^^ blue-shifted''
De Sitter Universes (respectively) \cite{Hel:The}. Both of them are
possesing a 10-dimensional group of motion (the De Sitter group) with
a globally defined timelike Killing vector, which is a very important
feature for defining an universal cosmic time, fulfiling this way the
Perfect Cosmological Principle \cite{Bar:The}, and moreover the
^^ ^^ red-shifted'' one contains also an event horizon
\cite{Hel:The,Wei-Gra}. At the present observational level it can
easily be shown that the ^^ ^^ blue-shifted'' De Sitter solution is
physically unacceptable \cite{Hel:The} and thus
\begin{equation}
f_{+} (t) = \sqrt{\frac{\Lambda}{3}} \, t \, , \; t \in
(- \infty , \, \infty)
\end{equation}
is the only physically significant solution of the
^^ ^^ energy'' - equation (16.b). For what it follows a short comment
on the exact dynamical role of each of the field equations (16) is
welcome. This regards the question ^^ ^^ what it would be if the
Universe were driven only by the
^^ ^^ cosmological pressure'' - equation ? ''
Mathematically the answer can be given immediately since, performing
in (16.a) the transformation
\begin{equation}
f = \alpha \ln u , \; \alpha \in R
\end{equation}
one gets the equation
\begin{equation}
u \left[ u,_{44} - \frac{\Lambda}{2 \alpha} u \right] + \left(
\frac{3}{2} \alpha - 1 \right) \left( u,_{4} \right)^{2} = 0
\end{equation}
and setting $\alpha = 2/3$ it results, omitting the trivial solution
$u=0$,
\begin{equation}
u,_{44} - \frac{3 \Lambda}{4} u = 0 .
\end{equation}
The general solution of (23) reads
\begin{equation}
u(t) = u_{+}^{0} \exp \left( \frac{\sqrt{3 \Lambda}}{2} t \right)
+ u_{-}^{0} \exp \left( - \frac{\sqrt{3 \Lambda}}{2} t \right) ,
\end{equation}
that means
\begin{equation}
f(t) = \frac{2}{3} \ln \left[ u_{+}^{0} \exp \left(
\frac{\sqrt{3 \Lambda}}{2} t \right) + u_{-}^{0} \exp \left(
- \frac{\sqrt{3 \Lambda}}{2} t \right) \right]
\end{equation}
and it represents a 2 - parameter family of spatially flat FRW
Universes driven by the ^^ ^^ cosmological pressure'' alone.
Physically, the result (25) combined with the
^^ ^^ energy'' - equation (16.b) points out two important features:
\begin{enumerate}
\item for $t \rightarrow \pm \infty$ the scale function behaves as
$f_{\pm}$ i.e. in the ^^ ^^ very distant'' past the 2 - parameter
family was degenerated in a ^^ ^^ blue-shifted'' De Sitter Universe
and will be degenerated again, this time in a ^^ ^^ red - shifted''
one, in the ^^ ^^ very distant'' future. Thus, for arbitrary
$u_{+}^{0} , \, u_{-}^{0}$ the pressure driven FRW - Universes
^^ ^^ have started'' and ^^ ^^ end'' as De Sitter spacetimes.
\item The general solution (25) of (16.a) naturally contains the De
Sitter solutions (18) of (16.b) for $u_{-}^{0} \equiv 0$ and
respectively $u_{+}^{0} \equiv 0$. Thus, putting it in other words,
the dynamics encoded in the ^^ ^^ cosmological pressure'' - equation
is more general than that controled by the ^^ ^^ energy'' - equation
but it might fail to be a physical one since (in general) it violates
the energy-spacetime geometry balance stated by (16.b).
\end{enumerate}

Generalizing this comment to the case when somehow the quantization
has been engaged it becomes clear that working with a field operator
related to $f$ by a suitable construction, one might solve the
quantum analogous of the ^^ ^^ cosmological pressure'' - equation,
expresses the Hamiltonian with respect to the derived solution for
the field operator and imposing at the end the quantum version of the
^^ ^^ energy'' - equation as a constraint on the formally possible
states of the system associated to its quantum dynamics. That's
exactly what we are going to do in the followings. Therefore, let's
come back to the action functional (6) and perform a partial
integration of the term containing the second derivative of the
metric function $f$ in order to get a ^^ ^^ surface'' - term and a
suitable form for the Lagrange function $L \left( f ; f,_{4}
\right)$ :
\begin{equation}
S[f] = \frac{3}{\kappa_{0}} \left[ \int d^{3} x \right]
\left[ e^{3f} f,_{4} \right]_{t_{-}}^{t_{+}} - \frac{3}{\kappa_{0}}
\left[ \int d^{3} x \right] \cdot \int e^{3f} \left[ \left( f,_{4}
\right)^{2} + \frac{\Lambda}{3} \right] dt
\end{equation}
Considering the spatial integration performed over an arbitrary
spacelike compact region of Euclidian volume $V$, (26) becomes
\begin{equation}
S[f] = \frac{V}{\kappa_{0}}
\left[ ( e^{3f} ),_{4} \right]_{t_{-}}^{t_{+}}
+ \int L \left(f ; f,_{4} \right) dt
\end{equation}
where
\begin{equation}
L(f ; f,_{4} ) = - \frac{3 V}{\kappa_{0}} e^{3f}
\left[ (f,_{4} )^{2} + \frac{\Lambda}{3} \right]
\end{equation}
is the Lagrangian of the spatially flat FRW Universe characterized by
the metric function $f$ that becomes a canonical variable. In terms
of $(f ; f,_{4})$ the Hamiltonian \[H[f] =
\frac{\partial L}{\partial f,_{4}} (f,_{4} ) - L \]
reads \[H[f] = - \frac{3V}{\kappa_{0}} e^{3f} \left[
(f,_{4} )^{2} - \frac{\Lambda}{3} \right] \]
and is quite obvious that the ^^ ^^ energy'' - equation (16.b) gets
the physically meaningfull expression \[H[f] = 0 \] that represents
the constraint imposed on the general solution of the Euler-Lagrange
equation (16.a) in order to be a physically acceptable one from a
classical viewpoint.

\section{Quantizing the Model}
\subsection{Standard Procedures}

Proceeding to quantization we notice that the substitution
\begin{equation}
\phi = \sqrt{\frac{8V}{3 \kappa_{0}}} \exp \left[ \frac{3}{2} f
\right]
\end{equation}
casts the Lagrangian (28) into the form
\begin{equation}
L[ \phi] = - \frac{1}{2} \left[ (\phi,_{4})^{2} + \frac{3 \Lambda}{4}
\phi^{2} \right]
\end{equation}
which, except the sign of the first term, is practically the same
with the Lagrangian density of a real scalar field of mass
$(3 \Lambda /4)^{1/2}$. This leads straigheforwardly to the
non-perturbative canonical quantization of the field $\phi$ since,
treating $\phi$ as a coordinate-like operator and
\begin{equation}
\pi = \frac{\partial L}{\partial \phi,_{4}} = - \phi,_{4}
\end{equation}
as its canonically conjugated momentum-operator, it is natural to
impose the Bohr-Heisenberg commutation relation
\begin{equation}
[ \pi , \phi] = - i.
\end{equation}
With (30) and(31), the associated Hamiltonian reads
\begin{equation}
H[\pi , \phi] = - \frac{1}{2} [\pi^{2} - \mu^{2} \phi^{2} ] ,
\end{equation}
where
\begin{equation}
\mu = \frac{\sqrt{3 \Lambda}}{2} ,
\end{equation}
Coming back to (30), we get the Euler-Lagrange equation as
\begin{equation}
\phi,_{44} - \mu^{2} \phi = 0
\end{equation}
and it yields for the field operator the hyperbolic expression
\begin{equation}
\phi = a_{+} e^{\mu t} + a_{-} e^{- \mu t}
\end{equation}
where, in order to do not violate at the classical level the reality
of $\phi$ , the operators $a_{\pm}$ have to be hermitians. With (36),
the commutation relation becomes
\[ [ a_{+} e^{\mu t} - a_{-} e^{- \mu t} , \,
a_{+}e^{\mu t} + a_{-} e^{- \mu t}] = \frac{i}{\mu} \]
i.e. the $a_{\pm}$ - hermitic operators must satisfy
\begin{equation}
[ a_{+} , \, a_{-} ] = \frac{i}{2 \mu}
\end{equation}
For later use, the Hamilton operator (33) can be worked out by (36)
in terms of $a_{\pm}$ :
\begin{equation}
H[\phi] = \mu^{2} [ a_{+} a_{-} + \, a_{-} a_{+}]
\end{equation}
Inspired by (37), (38) and their analogous formulas for the case of
the quantum harmonic oscillator, let us express the operators
$a_{\pm}$ as a linear combination with respect to a linear operator
^^ ^^ $C$'' and its adjoint ^^ ^^ $C^{+}$'' :
\begin{eqnarray}
& a) & a_{+} =  \frac{\rho}{\sqrt{2 \mu}} [e^{i \alpha} C
+ e^{-i \alpha} C^{+} ]\nonumber \\
& & \\ & b) & a_{-} =  \frac{\rho}{\sqrt{2 \mu}} [e^{i \beta} C
+ e^{-i \beta} C^{+}] \nonumber
\end{eqnarray}
where $\rho, \alpha, \beta \in R$. With (39) the commutation
relation (37) becomes \[\rho^{2} [ e^{i \alpha} C + e^{-i \alpha}
C^{+} , \, e^{i \beta} C + e^{-i \beta} C^{+} ] = i \] i.e. for
\begin{equation}
2 \rho^{2} \sin(\alpha - \beta) = 1 , \; {\rm with \, the \,
constraint \,} \alpha - \beta
\in \bigcup_{n} (2 n \pi , (2n+1) \pi ), n= 0,1,2,...
\end{equation}
it reads
\begin{equation}
[C, \, C^{+} ] = 1
\end{equation}
and it can be considered as the commutator between the annihilation
and creation operators $C, \, C^{+}$ respectively, which act on a
Fock-state of $n$ - quanta according to the formulas
\begin{eqnarray}
C \mid n \rangle = \sqrt{n} \mid n-1 \rangle  & ; & \langle n
\mid C^{+} = \sqrt{n} \langle n-1 \mid \nonumber \\  & & \\
C^{+} \mid n \rangle = \sqrt{n+1} \mid n+1 \rangle & ; &  \langle n
\mid C = \sqrt{n+1} \langle n+1 \mid    \nonumber
\end{eqnarray}
with
\begin{eqnarray*}
  C \mid 0 \rangle & = & 0 \\  \langle 0 \mid C^{+} & = & 0
\end{eqnarray*}
In terms of $C, \, C^{+}$, using the linear decomposition (39), the
Hamiltonian (38) gets the expression
\begin{equation}
H[\phi] = \mu \rho^{2} \left[ e^{i(\alpha + \beta)} (C)^{2} +
\cos(\alpha - \beta) (C C^{+} + C^{+} C ) + e^{-i(\alpha + \beta)}
(C^{+})^{2} \right]
\end{equation}
which leads by the natural condition
\begin{equation}
\langle 0 \mid H[\phi] \mid 0 \rangle = 0 ,
\end{equation}
using (42), to the equation
\begin{equation}
\cos(\alpha - \beta) = 0
\end{equation}
having, because of the constraint mentioned in (40), the solution
\begin{equation}
\alpha - \beta = (4k+1) \frac{\pi}{2} , \; k= 0,1,2,...
\end{equation}
Thus,\[ \sin(\alpha - \beta) = 1 , \; \rho^{2} = \frac{1}{2} , \;
\alpha = \beta + \frac{\pi}{2} + 2k \pi \] and, without restraining
the generality, we choose
\begin{eqnarray}
& a)  & \rho  = \frac{1}{\sqrt{2}} \nonumber \\ & &  \\
& b)   & \alpha  = \beta + \frac{\pi}{2} \nonumber
\end{eqnarray}
that casts (39) into the form
\begin{eqnarray}
& a) & a_{+}  =  \frac{i/2}{\sqrt{\mu}} [ e^{i \beta} C
- e^{-i \beta} C^{+} ] \nonumber \\ & &  \\
& b) & a_{-}  =  \frac{1/2}{\sqrt{\mu}} [ e^{i \beta} C
+ e^{-i \beta} C^{+} ]  \nonumber
\end{eqnarray}
and expresses the Hamiltonian (43) as:
\begin{equation}
H[\phi] = \frac{i \mu}{2} \left[ e^{2i \beta} (C)^{2}
- e^{-2i \beta} (C^{+})^{2} \right]
\end{equation}

With respect to the orthonormal basis $\lbrace \mid n \rangle ; \,
n=0,1,2,... \rbrace$ formed by the Fock-states of $n$ - quanta, the
Hamiltonian (49) is an infinite-dimensional extra-bidiagonal
Hermitian matrix with the elements
\begin{equation}
H_{n' n} = \frac{i \mu}{2} \left[ e^{2i \beta} \sqrt{n(n-1)} \,
\delta_{n' , n-2} - e^{-2i \beta} \sqrt{(n+1)(n+2)} \,
\delta_{n' , n+2} \right]
\end{equation}
and its eigenvalues are given by the ^^ ^^ secular'' - equation
\begin{equation}
{\rm det} \left[ \lambda \delta_{n' n} - \frac{i \mu}{2} \left(
e^{2i \beta} \sqrt{n(n-1)} \delta_{n' , n-2} - e^{-2i \beta}
\sqrt{(n+1)(n+2)} \delta_{n' , n+2} \right) \right] = 0
\end{equation}
As it can be easily seen, because of the Hermitian character of (50),
the phase $\beta$ doesn't affect the eigenvalues $\lambda$ (obtained
from (51)) and so it can be fixed arbitrarily. As an example, in the
case of a 9-th order truncation of the Hamiltonian matrix (50), its
eigenvalues $\lambda$, obtained by solving the corresponding 9-th
degree algebraic equation (51), are:
\begin{eqnarray}
a) & \lambda_{-4} & =   - \frac{\mu}{\sqrt{2}} \lbrace 22
+ \frac{3}{\sqrt{2}} [ 1 + (99 - 4 \sqrt{2})^{1/2} ] \rbrace^{1/2}
\nonumber \\
b) & \lambda_{-3} & =   - \frac{\mu}{\sqrt{2}} \lbrace 22
+ \frac{3}{\sqrt{2}} [ 1 - (99 - 4 \sqrt{2})^{1/2} ] \rbrace^{1/2}
\nonumber \\
c) & \lambda_{-2} & =  - \frac{\mu}{\sqrt{2}} [ 17
+ \sqrt{226}]^{1/2} \nonumber \\
d) & \lambda_{-1} & =  - \frac{\mu}{\sqrt{2}} [ 17
- \sqrt{226}]^{1/2} \nonumber \\
e) & \lambda_{0} & =  0 \\
f) & \lambda_{1} & =   \frac{\mu}{\sqrt{2}} [ 17 - \sqrt{226}]^{1/2}
\nonumber \\
g) & \lambda_{2} & =   \frac{\mu}{\sqrt{2}} [ 17 + \sqrt{226}]^{1/2}
\nonumber \\
h) & \lambda_{3} & =  \frac{\mu}{\sqrt{2}} \lbrace 22
+ \frac{3}{\sqrt{2}} [ 1 - (99 - 4 \sqrt{2} )^{1/2}] \rbrace^{1/2}
\nonumber \\
i) & \lambda_{4} & =  \frac{\mu}{\sqrt{2}} \lbrace 22
+ \frac{3}{\sqrt{2}} [ 1 + (99 - 4 \sqrt{2} )^{1/2}] \rbrace^{1/2}
\nonumber
\end{eqnarray}
It can be noticed that the negative - and positive - energy
eigenstates of $H$ appear in pairs, but this is not a problem, in
what it concerns the negative - energy eigenstates, for the physical
states $\mid \psi \rangle $ have to safisfy the
^^ ^^ energy-geometry'' - constraint, i.e. Wheeler-DeWitt equation
\begin{equation}
H[ \phi] \mid \psi \rangle = 0
\end{equation}
which, for the analyzed system, is the quantum analogous of the
^^ ^^ energy'' - equation (16.b).

\subsection{Some Quantum Properties}

Before dealing with (53) in order to compute the physical solution(s)
$ \mid \psi \rangle $, it is useful to derive some of the quantum
properties ^^ ^^ generated'' by the field operator (36), with the
linear decomposition (48), and by the Hamilton operator (49).
\begin{itemize}
\item Starting with the field operator $\phi$ casted into the form
\begin{equation}
\phi = \frac{1}{\sqrt{\mu}} \left \lbrace \frac{i}{2}
[ e^{i \beta} C - e^{-i \beta} C^{+} ] e^{\mu t} + \frac{1}{2}
[ e^{i \beta} C + e^{-i \beta} C^{+} ] e^{- \mu t}  \right \rbrace
\end{equation}
is clear first that, as it should be, its mean value in a given
Fock-state of $n$ - quanta is
\begin{equation}
\langle n \mid \phi \mid n \rangle \equiv 0, \; \forall \, n=0,1,2,...
\end{equation}
For the Hermitic operator $\phi^{2}$ the situation is different and
we get
\begin{eqnarray*}
\phi^{2} = & - & \frac{1}{4 \mu}  [e^{2i \beta}
(C)^{2} - (C C^{+} + C^{+} C ) + e^{-2i \beta}
(C^{+})^{2} ] e^{2 \mu t}  +  \\
 & + & \frac{1}{4 \mu}  [e^{2i \beta}
(C)^{2} + (C C^{+} + C^{+} C ) + e^{-2i \beta}
(C^{+})^{2} ] e^{- 2 \mu t} +  \\
 &  + & \frac{i}{4 \mu}  [e^{2i \beta}
(C)^{2} - e^{-2i \beta} (C^{+})^{2}]
\end{eqnarray*}
i.e.
\begin{equation}
\langle n \mid \phi^{2} \mid n \rangle = \frac{n+ 1/2}{\mu}
\cosh(2 \mu t)
\end{equation}
which means that the mean-deviation of the field $\phi$, expressing
its quantum fluctuation at any given moment $t$ in a state with
$n$ - quanta, is given by the formula
\begin{equation}
(\triangle \phi)_{n} (t) = \left[ \frac{n+1/2}{\mu} \cosh(2 \mu t)
\right]^{1/2}
\end{equation}
Having a look at (29), which states the classical relation between
$\phi$ and the scale-function $f$, we also derive from (56) that
\begin{equation}
\langle n \mid e^{3f} \mid n \rangle = \frac{\kappa_{0}}{4V}
\sqrt{\frac{3}{\Lambda}} (n + 1/2) \cosh(\sqrt{3 \Lambda} t)
\end{equation}
As $V$ is the Euclidian volume of an (arbitrary) $t$ - leveled
compact region, it results that
\begin{equation}
\mho = \int _{(V)} \sqrt{-g} d^{3} x = V e^{3f}
\end{equation}
represents its ^^ ^^ universal'' - volume and so (58) it yields
\begin{equation}
\mho_{n} (t) = \frac{\kappa_{0}}{4} \sqrt{\frac{3}{\Lambda}}
( n+ 1/2) \cosh(\sqrt{3 \Lambda} t)
\end{equation}
pointing out a very interesting volume-quantization at any instant
$t$ in an $n$ - quanta state. Probably the most puzzling feature
outlined by (60) is that in comparison to the classical case when at
any instant $t$, $\mho$ can be arbitrary large (because of V), in the
quantum case the accesible universal-volume in a state of
$n$ - quanta (at the instant $t$) {\it is finite}. As a consequence,
at the Minkowskian epoch $t=0$ in the Fock state of 0 - quanta, it
exists an elementary 0-universal-volume
\[ \mho_{0} = \frac{\kappa_{0}}{8} \sqrt{\frac{3}{\Lambda}} \]
which, considering $\Lambda$ as a fundamental constant, is expressed
only with respect to the universal constants $c, \hbar, G, \Lambda$ :
\begin{equation}
\mho_{0} = \frac{\pi G \hbar}{c^{3}} \sqrt{\frac{3}{\Lambda}}
\end{equation}
\item
Passing to $H[\phi]$ given by (49) it results, according to (50)
also, that its mean-value in any $n$ - quanta state is zero,
\begin{equation}
\langle n \mid H[\phi] \mid n \rangle \equiv 0,
\end{equation}
and for the square-mean-deviation it yields
\[(\triangle H)_{n}^{2} = \langle n \mid H^{2} [\phi] \mid n \rangle
= \frac{\mu^{2}}{4} \langle n \mid
(C)^{2} (C^{+})^{2} + (C^{+})^{2} (C)^{2} \mid n \rangle , \]  i.e.
\begin{equation}
\langle n \mid H^{2} [\phi] \mid n \rangle
= \frac{\mu^{2}}{2} (n^{2} + n + 1)
\end{equation}
and so, the mean-fluctuation of $H$ in the Fock-state
$\mid n \rangle $ reads
\begin{equation}
(\triangle H)_{n} = \frac{\mu}{\sqrt{2}} (n^{2} + n + 1)^{1/2} ,
\end{equation}
increasing linearly with $n$ at large values. Therefore the
^^ ^^ energy per quanta''- mean-fluctuation is approximately constant
\begin{equation}
\frac{(\triangle H)_{n}}{n} \cong \frac{\mu}{\sqrt{2}}
\end{equation}
being equal with the energy-mean-fluctuation in the
^^ ^^ no-quanta'' - state, $\mid 0 \rangle $. This might be an
expression of the ^^ ^^ time-energy'' Heisenberg relation because in
a continuesly evolving spatially flat FRW Universe (with cosmological
constant) the Fock-states
$\lbrace \mid n \rangle \rbrace_{n=0,1,2,...}$
are no longer invariant under the action of $H$,
\begin{equation}
H \mid n \rangle = \frac{i \mu}{2} \lbrace e^{2i \beta}
\sqrt{n(n-1)} \mid n-2 \rangle - e^{- 2i \beta} \sqrt{(n+1)(n+2)}
\mid n+2 \rangle \rbrace ,
\end{equation}
hence the number of ^^ ^^ FRW - quanta'' is no longer conserved, and
so, the mean-life-time $\tau$ of a ^^ ^^ FRW - quanta'' can be
estimated from
\begin{equation}
\frac{(\triangle H)_{n}}{n} \cdot \tau \approx 1 \; \Leftrightarrow
\; \frac{\mu}{\sqrt{2}} \tau \approx 1
\end{equation}
i.e.
\begin{equation}
\tau = \sqrt{\frac{8/3}{\Lambda}}
\end{equation}
\end{itemize}

By the way, in what it concerns the quanta-number non-conservation,
let us derive the differential equation satisfied by the
number-operator
\begin{equation}
N = C^{+} C
\end{equation}
and work out the ^^ ^^ phenomenological'' solution $n(t)$. First, we
have
\[ \frac{dN}{dt} = i [ H,N] = - \frac{\mu}{2} \lbrace e^{2i \beta}
[ C^{2} , C^{+} C ] - e^{-2i \beta} [(C^{+})^{2} , C^{+} C] \rbrace
\] i.e.
\begin{equation}
\frac{dN}{dt} = - \mu \lbrace e^{2i \beta} (C)^{2} + e^{2i \beta}
(C^{+})^{2} \rbrace
\end{equation}
For the second-order derivative of the number-operator it results
\[ \frac{d^{2}N}{dt^{2}} = i \left[ H , \frac{dN}{dt} \right]
= \mu^{2} [(C)^{2} , (C^{+})^{2} ], \]
such that, the differential equation obeyed by $N$ reads:
\begin{equation}
\frac{d^{2}N}{dt^{2}} - 4 \mu^{2} N = 2 \mu^{2}
\end{equation}
Its averaged-version,
\begin{equation}
\frac{d^{2}n}{dt^{2}} - 4 \mu^{2} n = 2 \mu^{2} \, ,
\end{equation}
rules the quanta-number dynamics, $n(t)$, possessing the general
solution
\begin{equation}
n(t) = n_{1}^{0} e^{2 \mu t} + n_{2}^{0} e^{-2 \mu t} - 1/2
\end{equation}
and the phenomenological ^^ ^^ annihilation - creation rate'' of
FRW-quanta reads
\begin{equation}
S(t) = \frac{dn}{dt} = 2 \mu [n_{1}^{0} e^{2 \mu t} - n_{2}^{0}
e^{-2 \mu t}] \end{equation}
Imposing $n_{1}^{0} , n_{2}^{0} > 0$ as a necessary condition for
$n(t) \geq 0, \forall t \in (- \infty , \infty)$ the relations (73),
(74) point out physically that in the ^^ ^^ very distant'' past,
$t \rightarrow - \infty$, the Universe was a
De Sitter - ^^ ^^ blue-shifted'' one filled with a very large number
of quanta \[ n_{- \infty} (t) \cong n_{2}^{0} e^{-2 \mu t} \]
possesing a huge phenomenological absorbtion rate
\[  S_{- \infty} (t) \cong -2 \mu n_{2}^{0} e^{-2 \mu t} \]
As the time passes everything slows down but the phenomenological
creation rate is getting bigger and bigger and after some stationary
epoch $t_{0}$ (depending on $n_{1}^{0}, n_{2}^{0}$) the Universe
becomes a De Sitter - ^^ ^^ red-shifted'' one with a very large
number of quanta \[ n_{+ \infty} (t) \cong  n_{1}^{0} e^{2 \mu t} \]
and a huge pnenomenological creation rate
\[S_{+\infty} (t) \cong 2 \mu n_{1}^{0} e^{2 \mu t} \]
for compensating the exponential rate $\mho \sim e^{2 \mu t}$ of the
universal-volume expantion.

\subsection{The Physical Quantum States}

Finally, let us come back to the Wheeler-DeWitt equation (53) which,
inserting (49), concretely reads:
\begin{equation}
\left[ e^{i(2 \beta + \frac{\pi}{2})} (C)^{2} + e^{-i (2 \beta
+ \frac{\pi}{2})} (C^{+})^{2} \right] \mid \psi > \, = \, 0
\end{equation}
As the phase $\beta$ is dynamically irrelevant, let us set
\begin{equation}
\beta = - \frac{\pi}{4}
\end{equation}
dealing this way with the ^^ ^^ very symmetric'' equation:
\begin{equation}
[(C^{+})^{2} + (C)^{2} ] \mid \psi > \, = \, 0.
\end{equation}
Expressing $ \mid \psi \rangle $ as a mixed quantum state with
respect to the orthonormal Fock-basis
$ \lbrace \mid n \rangle  \rbrace_{n=0,1,2,...}$ i.e.
\begin{equation}
\mid \psi \rangle  = \sum_{n=0}^{\infty} a_{n} \mid n \rangle
\end{equation}
we get from (78) the state equation
\begin{eqnarray}
& & \sqrt{2} a_{2} \mid 0 \rangle  +  \sqrt{2 \cdot 3} a_{3} \mid 1
\rangle + \nonumber \\ & &  \\
& & + \sum_{n=2}^{\infty} \left[ \sqrt{(n+1)(n+2)} a_{n+2}
+  \sqrt{n(n-1)} a_{n-2} \right] \mid n \rangle = 0 \nonumber
\end{eqnarray}
that demands
\begin{equation}
a_{2} = a_{3} \equiv 0
\end{equation}
and transforms into the difference-equation
\begin{equation}
\sqrt{(n+1)(n+2)} \, a_{n+2} + \sqrt{n(n-1)} \, a_{n-2}  = 0, \;
n=2,3,...
\end{equation}
that must be safisfied by the amplitudes
$\left \lbrace a_{n} \right \rbrace_{n=0,1,2,...}$
of the physical quantum state(s) of spatially-flat FRW-Universe with
cosmological constant, $\Lambda$. A very interesting feature of (81)
with the necessary and sufficient condition (80) (for admitting a
non-trivial solution) is that there are two fundamental decoupled
states, i.e. orthonormal,
$ \mid \psi_{+} \rangle , \; \mid \psi_{-} \rangle $
possesing the general forms
\begin{eqnarray}
& a) & \mid \psi_{+} \rangle  =  \sum_{k=0}^{\infty} a_{4k} \mid 4k
\rangle \nonumber \\
& & \\ & b) & \mid \psi_{-} \rangle  =  \sum_{k=0}^{\infty} a_{4k+1}
\mid 4k +1 \rangle \nonumber
\end{eqnarray}
such that ^^ ^^ $ \psi_{+}$'' - , \, ^^ ^^ $\psi_{-}$'' - difference
equations obtained from (81) explicitly read
\begin{eqnarray}
 & a) & [(4k+1)(4k+2)]^{1/2} a_{4k+2} + [4k(4k-1)]^{1/2} a_{4k-2}
=  0  \nonumber \\  & & \\
 & b)  &  [(4k+2)(4k+3)]^{1/2} a_{4k+3} + [4k(4k+1)]^{1/2} a_{4k-1}
=  0 \nonumber
\end{eqnarray}
the non-trivial solutions being obtained, because of (80), for
$k= \frac{1}{2} , \frac{3}{2} , \frac{5}{2} , ...$

In order not to deal with semi-integer values for $k$ and also in
order to work directly with the ^^ ^^ index-structure'' of (82) the
rescaling $4k \rightarrow 4 k + 2$ in both (83.a) and (83.b) is
extremely convenient because it separates at once the
recurrent-equations
\begin{eqnarray}
 a) &  a_{4k+4} = - \left \lbrace \frac{(4k+1)(4k+2)}{(4k+3)(4k+4)}
\right \rbrace^{1/2} a_{4k}  &  \nonumber \\
 & & ;  k= 0,1,2,... \\
 b) & \; \;  a_{4k+5} = - \left \lbrace
\frac{(4k+2)(4k+3)}{(4k+4)(4k+5)} \right \rbrace^{1/2} a_{4k+1} &
  \nonumber
\end{eqnarray}
for the physical amplitudes
$\left \lbrace a_{4k} \right \rbrace_{k \in \rm{N}} ,\left
\lbrace a_{4k+1} \right \rbrace_{k \in \rm{N}}$
of the states $ \mid \psi_{+} \rangle, \mid \psi_{-} \rangle$
given by (82). Hence, the solutions of (84.a), (84.b) read
respectively
\begin{eqnarray}
a) \; \; \; \;  a_{4n} = (-1)^{n} \prod_{k=0}^{n-1} & \left \lbrace
\left[ \frac{(4k+1)(4k+2)}{(4k+3)(4k+4)} \right]^{1/2} \right \rbrace
a_{0} &   \nonumber \\  & & ; \; n= 1,2,... \\
b)  \; \; a_{4n+1} = (-1)^{n} \prod_{k=0}^{n-1} & \left \lbrace
\left[ \frac{(4k+2)(4k+3)}{(4k+4)(4k+5)} \right]^{1/2} \right \rbrace
a_{1} &  \nonumber
\end{eqnarray}
and the amplitudes $a_{0} , a_{1}$ are fixed up to a phase factor by
the normalization
\begin{eqnarray}
& a) & \mid a_{0} \mid =  \left \lbrace 1 + \sum_{n=1}^{\infty}  \mid
\frac{a_{4n}}{a_{0}}  \mid^{2} \right \rbrace^{-1/2} =
\left \lbrace \sum_{n=0}^{\infty} \frac{(2n)!}{2^{2n} (n!)^{2}}
\frac{\Gamma(3/4)}{\Gamma(1/4)} \frac{\Gamma(n+ 1/4)}{ \Gamma(n+3/4)}
\right \rbrace^{-1/2}
\nonumber \\ & & \\ & b)
& \mid a_{1} \mid =  \left \lbrace 1 + \sum_{n=1}^{\infty}  \mid
\frac{a_{4n+1}}{a_{1}}  \mid^{2} \right \rbrace^{-1/2} = \left
\lbrace \sum_{n=0}^{\infty} \frac{(2n)!}{2^{2n} (n!)^{2}}
\frac{\Gamma(5/4)}{\Gamma(3/4)} \frac{\Gamma(n+ 3/4)}{ \Gamma(n+5/4)}
\right \rbrace^{-1/2} \nonumber
\end{eqnarray}

In the end, as $ \mid \psi_{+} \rangle , \mid \psi_{-} \rangle$ given
by (82), with (85), (86), are two physical quantum states of the
system, let us compute, at least formally, the effective propagators
\begin{eqnarray}
& a) & G_{+} (t,t_{0}) = \langle \psi_{+} \mid {\rm T} [ \phi(t)
\phi(t_{0} )] \mid \psi_{+} \rangle  \nonumber \\ & & \\
& b) & G_{-} (t,t_{0}) = \langle \psi_{-} \mid {\rm T} [ \phi(t)
\phi(t_{0}) ] \mid \psi_{-} \rangle,  \nonumber
\end{eqnarray}
where the ${\rm T}$ - operation stands for the usual time-ordered
product. So, using the field operator expression (54), we get
\begin{eqnarray}
{\rm T} [\phi(t) \phi(t_{0})] =  \frac{1/2}{\mu}  \lbrace
[ C^{+} C + C C^{+} ] \cosh[\mu(t+t_{0})] + i \sinh [ \mu(t - t_{0})]
+   \nonumber \\    +  i [ e^{2 i \beta} (C)^{2} - e^{-2i \beta}
(C^{+})^{2} ] \cosh [ \mu(t-t_{0})] -  \\      -  [e^{2i \beta}
(C)^{2} + e^{-2i \beta} (C^{+})^{2} ] \sinh [\mu (t+t_{0} ) ]
\rbrace , \nonumber
\end{eqnarray}
with $t \geq t_{0}$, and because
\begin{eqnarray}
\langle \psi_{+} \mid (C)^{2} \mid \psi_{+} \rangle = 0 & , &
\langle \psi_{+} \mid (C^{+})^{2} \mid \psi_{+} \rangle = 0
\nonumber \\ & & \\ \langle \psi_{-} \mid (C)^{2} \mid \psi_{-}
\rangle = 0 & , & \langle \psi_{-} \mid (C^{+})^{2} \mid \psi_{-}
\rangle = 0  \nonumber
\end{eqnarray}
(as it can be easily seen in the light of (80)) it results from (88)
after some simple calculations :

\newpage

\begin{eqnarray}
a) & G_{+} (t, t_{0}) = & \frac{1/2}{\mu}  \lbrace
\cosh [ \mu (t+ t_{0})] + i \sinh[\mu (t-t_{0})] +  \nonumber \\
& &  + \; 8 \; \frac{\sum_{n=1}^{\infty} n  \frac{(2n)!}{2^{2n}
(n!)^{2}} \frac{\Gamma(n+ 1/4)}{ \Gamma(n+3/4)}}{\sum_{n=0}^{\infty}
\frac{(2n)!}{2^{2n} (n!)^{2}} \frac{\Gamma(n+ 1/4)}{ \Gamma(n+3/4)}}
\cosh [ \mu (t+ t_{0})]    \rbrace  \nonumber   \\ & &  \\
b)   & G_{-} (t, t_{0}) = & \frac{1/2}{\mu}  \lbrace
3 \cosh [ \mu (t+ t_{0})] + i \sinh[\mu (t-t_{0})] +  \nonumber \\
& &  + \; 8 \; \frac{\sum_{n=1}^{\infty} n  \frac{(2n)!}{2^{2n}
(n!)^{2}} \frac{\Gamma(n+ 3/4)}{ \Gamma(n+5/4)}}{\sum_{n=0}^{\infty}
\frac{(2n)!}{2^{2n} (n!)^{2}} \frac{\Gamma(n+ 3/4)}{ \Gamma(n+5/4)}}
\cosh [ \mu (t+ t_{0})]    \rbrace  \nonumber
\end{eqnarray}

\section{Summary and Discussion}

We have noticed that, with respect to the modified coherent metric
field $\phi(t)$, the spatially-flat Robertson-Walker Universe with
cosmological constant is an exactly solvable Hamiltonian system, with
the Lagrangian (30) of the ^^ ^^ reversed oscillator'' - type. This
feature allows a canonical non-perturbative quantization under the
Wheeler - DeWitt constraint (53), which arises naturally as the
quantum analogous of the Einsteinian ^^ ^^ energy'' - equation (16.b)
and filters the two orthogonal physical states (82) of the (quantum)
system, exactly as Einstein equation did at the classical level, with
the disjoint ^^ ^^ blue'' and ^^ ^^ red - shifted'' De Sitter (exact)
solutions. The derived expression of the field operator (54) casting
the Hamiltonian into the form (43) leads to the non-conservation of
the number of ^^ ^^ FRW - quanta'', mathematically viewed as the
eigenvalues of the number operator. Therefore, each $n$ - quanta Fock
state is non-trivially evolving, exhibiting energy fluctuations and a
finite ^^ ^^ universal'' - volume quantization, (60), induced by the
modified coherent metric field fluctuations. Spontaneously, at
$t_{0} = t$, the effective field propagators (90) are real and point
out that the large scale infinity of the (modeled) Universe is due to
the quantum - geometrical contributions of \underline{all}
$\mid 4n \rangle$ or $\mid 4n+1 \rangle$ ^^ ^^ FRW - quanta'' that
build up its physical states $\mid \psi_{\pm} \rangle$.  \\ \\

\newpage

{\large{\bf Acknowledgments}} \\ \\
One of us (C.D.) wishes to express his deepest gratitude to the
Japanese Government for financially supporting his work under a
MONBUSHO Fellowship. The kind hospitality of the Toyama University is
highly appreciated by (M.A.D.) and (C.D.). We are also greatly
indebted to the Quantum Theory Group from the Physics Department of
Toyama University for the stimulating scientific environment while
this work was carried out.

\end{document}